# Preprint notes

**Title of the article:**

Implementing AI Ethics in Practice: An Empirical Evaluation of the RESOLVEDD Strategy

**Authors:**

Ville Vakkuri and Kai-Kristian Kemell

**Notes:**

- This is the author's version of the work
- The definite version was published in: Vakkuri V., Kemell KK. (2019) Implementing AI Ethics in Practice: An Empirical Evaluation of the RESOLVEDD Strategy. In: Hyrynsalmi S., Suoranta M., Nguyen-Duc A., Tyrväinen P., Abrahamsson P. (eds) Software Business. ICSOB 2019. Lecture Notes in Business Information Processing, vol 370. Springer, Cham
- Copyright owner's version can be accessed at DOI: https://doi.org/10.1007/978-3-030-33742-1_21

# Implementing AI Ethics in Practice: An Empirical Evaluation of the RESOLVEDD Strategy[1]

**Ville Vakkuri**[0000-0002-1550-1110], **Kai-Kristian Kemell**[0000-0002-0225-4560]

Faculty of Information Technology, University of Jyväskylä
Jyväskylä, Finland
ville.vakkuri@jyu.fi, kai-kristian.o.kemell@jyu.fi

*Abstract—* As Artificial Intelligence (AI) systems exert a growing influence on society, real-life incidents begin to underline the importance of AI Ethics. Though calls for more ethical AI systems have been voiced by scholars and the general public alike, few empirical studies on the topic exist. Similarly, few tools and methods designed for implementing AI ethics into practice currently exist. To provide empirical data into this on-going discussion, we empirically evaluate an existing method from the field of business ethics, the RESOLVEDD strategy, in the context of ethical system development. We evaluated RESOLVEDD by means of a multiple case study of five student projects where its use was given as one of the design requirements for the projects. One of our key findings is that, even though the use of the ethical method was forced upon the participants, its utilization nonetheless facilitated of ethical consideration in the projects. Specifically, it resulted in the developers displaying more responsibility, even though the use of the tool did not stem from intrinsic motivation.

## 1. Introduction

As Artificial Intelligence (AI) and Autonomous Systems (AS) become increasingly ubiquitous, real-life incidents, such as the recent Cambridge Analytica one, begin to highlight the importance of AI ethics. AI systems are unique in that one cannot opt out of using them. Even if one does not own an autonomous vehicle, it would seem that one nonetheless has to drive on the roads with them. Similarly, one cannot avoid being tracked by AI-based surveillance systems even if one does not consent to being surveilled. In this fashion, the very idea of an active user in the context of AI systems becomes blurred as human actors, e.g. become mere objects of data collection.

As the enormous impact of AI systems becomes increasingly clear, calls for privacy and fairness in these systems grow more prominent. The city of San Francisco already voted to ban facial recognition from being used to track and profile its citizens[2], underlining that regulations and laws directed at AI systems are likely to grow in number as further progress on AI is made. With laws and regulations (e.g. GDPR) starting to necessitate ethical consideration in AI design, and with the general public demanding more ethical systems, those utilizing or designing these systems should become familiar with AI ethics.

Organizations developing and deploying AI systems will arguably benefit from focusing on fair systems that respect the privacy of their users in the future. With such trends as environmental awareness and user privacy, ethics seem to be becoming a global mega trend. As users become more aware of their privacy and how data is handled by various AI systems, ethical development is likely to become a selling point for such systems.

Studies in the area of AI ethics should seek to bridge this gap between research and practice by turning to the field of behavioral Software Engineering (SE) [18]. If the goal is to make ethics a part of AI system development, the focus should be on the developers. In practice, it is the developers who build the ethical principles into the system, as no AI system is at present capable of evaluating and deciding on its own ethical principles. In doing so, developers build their own values into the systems, which end up reflecting their views [2]. Yet, it is known that developers are not well-informed of ethics in software engineering [19].

This, combined with the current lack of tools and methods in AI ethics, has resulted in a situation where developers do not have the means to implement ethics. The methods that exist have not seen widespread adoption [26] and lack empirical validation or are immature [21]. In developing methods for this area, the focus should be on understanding the developers, focusing on behavioral SE [18].

Currently, ethical issues seem to be often simplified or neglected entirely during development. This can be costly when they then later surface during the operational life of the system, as was e.g. the case when Amazon's recruitment

---

[1] An early version of this paper was presented in the Euromicro Conference on Software Engineering and Advanced Applications (SEAA 2019)

[2] https://www.bbc.com/news/technology-48276660

AI[3] was found to be biased towards women, having been trained using past recruitment data which featured predominantly male recruits.

Studies into implementing AI ethics in practice are currently lacking. Moreover, the methods and tools that we presently have are also lacking in empirical validation [21]. To provide empirical data into this area of research, in this paper we test an ethical tool from business ethics, the RESOLVEDD strategy[22], in the context of AI design. We do so by means of a multiple case study of five different prototype projects where the use of RESOLVEDD was one of the requirements for the projects. The goal of this study is to further our understanding on how to provide actionable tools for implementing AI ethics. In this paper, we approach this problem through the following two research questions:

1. Does the use of an ethical tool enhance ethical consideration in the design process?
2. How does the ethical tool RESOLVEDD perform in the AI context?

## 2. BACKGROUND

### A. Ethically Aligned Design

In the field of IT and ICT, ethics has historically been discussed in different contexts. It has been discussed in relation to (1) applying traditional ethical theories in the context of ICT; (2) as a branch of professional ethics for ICT; and (3) as a set of specific ethical issues such as internet privacy and security in ICT [5]. For example, traditional ethical theories such as Kantian ethics and virtue ethics have been applied in the context of ICT. Moreover, specific, practical questions related to professional ethics have been addressed in the ACM Code of Ethics [13]. In this paper, we define ethics from the point of view of ICT as follows: "the analysis of the nature and social impact of computer technology and the corresponding formulation and justification of policies for the ethical use of such technology" [20]. Another central construct used in this paper, Ethically Aligned Design [10], on the other hand refers to the involvement of decision-making in practice and ethical consideration in the practice and design AI and autonomous systems and technologies.

The continuing progress in the field of AI calls for new and concrete methods to manage the ethical issues arising from these new innovations [2,6]. Indeed, Allen et al. [2] argue that AI and AI-based systems produce new kinds of needs to consider. Specifically, they propose that designers implicitly embed values in the technologies they produce [2]. AI and other complex systems force designers to consider what kind of values are embedded in the technologies and also how the practical implementation of these values could be done and how these systems can be governed [6].

To better incorporate human values into the design process of AI systems, some AI-specific values have been proposed. For example, the importance of transparency in AI systems was emphasized by Bryson and Winfield [4]. Dignum [7] presented two more values in addition to transparency by presenting the ART principles (Accountability, Responsibility, Transparency) to guide ethical development of AI systems [7]. Finally, fairness and freedom from machine bias have also become important as core values expected from AI systems[12].

To direct the discussion on aligning ethics with system design, the IEEE Global Initiative on Ethics of Autonomous and Intelligent Systems was launched. The initiative was branded under a concept titled Ethically Aligned Design (EAD), a construct we briefly discussed at the start of this section. The initiative aims to encourage practitioners to consider and prioritize ethics in the development of AI. So far, the initiative has defined values and ethical principles that prioritize human well-being in a given cultural context. These guidelines have been published online (latest Edition1 2019).[10] These guidelines revolve around presenting different AI ethics issues and then suggesting ways of tackling each issue through extant literature, but ultimately offer very little in terms of actionable practices or tools, with most of the focus being on discussing the issues.

Arguably, the key audience of EAD are, or should be, the developers. AI development, much like conventional software development, is a cognitive activity [14] where humans play a significant role in deciding how the system behaves. Extant research has established that developers' interests are driven by work related concerns [1]. Concerns are the foundation of developer commitment development in his/her work. Commitment is important as it directs attention and helps in maintaining the chosen course of action [1]. Should EAD practices become used by the developers, it should be meaningful to them, contributing to their work related concerns and thus helping them accomplish their tasks.

Experiencing meaningfulness in the work place plays a significant role in understanding the ethical aspects related to one's work. Bowie [3] states that an overall experience of meaningfulness while working supports the individual's moral development related to that activity. Understanding the ethical aspects of one's work stems from understanding the meanings of one's own actions and responsibility for the well-being of others [3]. In this regard, the challenge in software and interactive systems development and design is that the developers may not fully understand the consequences of their actions and how their decisions eventually affect others once the system is operational. In other words, in order for EAD to be possible, ethics needs to become meaningful for developers. For ethics to become meaningful for developers, it needs to help developers accomplish work tasks, instead of being something extra they have to take into consideration, e.g. because the product manager tells them to.

---

[3] https://www.reuters.com/article/us-amazon-com-jobs-automation-insight/amazon-scraps-secret-ai-recruiting-tool-that-showed-bias-against-women-idUSKCN1MK08G

In summary, there are multiple methods that could potentially be used to implement AI ethics. However, we argue that AI calls for new, actionable methods to address the new ethical issues presented by these systems, specifically tailored for the context of AI. In the next section, we further discuss an existing tool for ethical decision-making that we focus on in this study, RESOLVEDD.

ethics [17]. Based on their study, Johansen [17] note that the method introduces a capability to produce a description of various solutions and viewpoints to a single problem. However, they also criticize the method for being time-consuming, and for giving no feedback to its users on whether they succeeded in implementing ethics. Indeed, as RESOLVEDD does not directly offer any solutions to the ethical issues it may help discover, it is up to its users how to address them, or whether to address them at all.

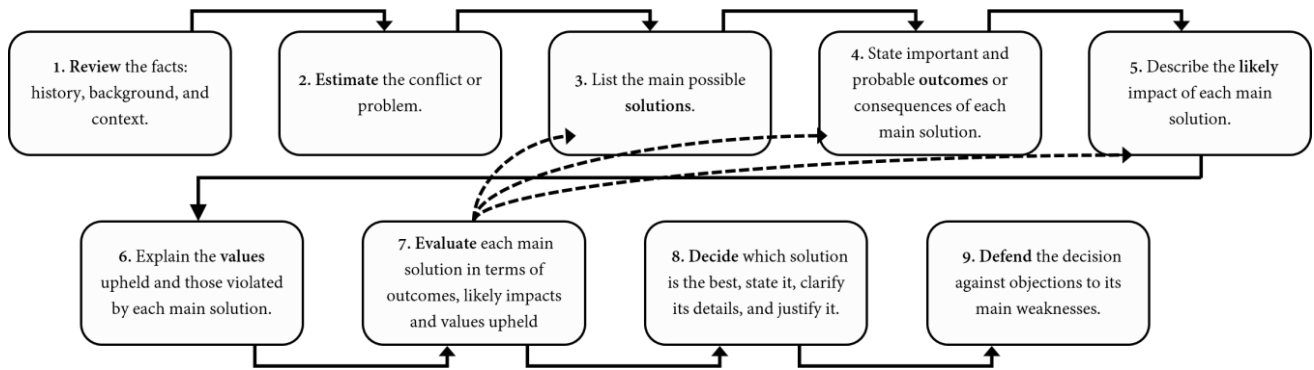

Figure 1 The Nine Steps of the RESOLVEDD Strategy

### B. The RESOLVEDD Strategy

The RESOLVEDD strategy was first introduced by Pfeiffer and Forsberg [22]. It is a step-by-step decision-making method, originally intended for teaching practical ethics to bachelor students. The method is aimed at those who do not have prior knowledge of ethics or philosophy to evaluate ethical principles in practice. This aspect of the RESOLVEDD strategy makes it particularly appealing in the field of Software Engineering (SE) where few curricula have traditionally included studies in ethics or philosophy.

The RESOLVEDD strategy is based on professional ethics and approaches ethics from the point of view of personal ethical problems in work contexts. It is not connected to any specific ethics theory and does not enforce any set of values on its would-be users. Instead, RESOLVEDD is intended to support its users in taking into account ethical issues and tackling them through their own set of values or through an ethics theory of their choice.[22]

The strategy is presented as a series of nine concrete steps (Figure 1) portraying the rational ethical decision-making process. By using the method, one is able to justify and explain the decision-making process leading up to whatever actions were ultimately taken. It is intended to help its users understand the ethical issues present in their work and encourages them to address them in the way they deem best, though nonetheless without compromising ethical principles. Though it originates from the field of business ethics, the method can also be utilized for tackling ethical issues outside the field of business. [22]

In extant research, the RESOLVEDD strategy has been applied in the field of biology where it was used to teach

### 3. RESEARCH MODEL

In addressing ethics as a part of AI development, various principles have been discussed in academic literature. For the time being, the discussion has centered on four constructs: Transparency [4,7,10], Accountability [7,10], Responsibility [7,10] and Fairness e.g.[12]. A recent EU report [11] also discussed Trustworthiness as a goal AI systems should strive for. Moreover, the field of AI ethics can be divided into three categories: (1) Ethics by Design (integration of ethical reasoning capabilities as a part of system behavior e.g. ethical robots); (2) Ethics in Design (the regulatory and engineering methods); and (3) Ethics for Design: (codes of conduct, standards etc.) [8]. In this paper, we focus on the ethically aligned development process.

Out of the aforementioned four main principles for AI Ethics, we consider accountability, responsibility, and transparency (the so-called ART principles, formulated by Dignum [7]) a starting point in understanding the involvement of ethics in AI projects. We have selected these three constructs as the basis of our research framework (Figure 2).

Transparency is defined in the ART principles of Dignum [7] as transparency of the AI systems, algorithms and data used, their provenance and their dynamics. I.e. transparency refers to understanding how AI systems work by being able to inspect them. Transparency can be argued to currently be the most important of these principles or values in AI ethics. Turilli and Floridi [25] argue that transparency is the key pro-ethical circumstance that makes it possible to implement AI ethics. It is also one of the key ethical principles in EAD [10].

In the research framework of this study, transparency is considered on two levels: (a) transparency of data and algorithms, as well as (b) transparency of systems development. The former refers to understanding the inner workings of the system in a given situation, while the latter refers to understanding what decisions were made by whom during development. It is a pro-ethical circumstance that makes it possible to assess accountability and responsibility.

Accountability refers to determining who is accountable or liable for the decisions made by the AI. Dignum [7] defines accountability to be the explanation and justification of one's decisions and actions to the relevant stakeholders. Transparency is required for accountability, as we must understand why the system acts in a certain fashion, as well as who made what decisions during development in order to establish accountability. Whereas accountability can be considered to be externally motivated, closely related but separate construct responsibility is internally motivated. In the context of this research framework, accountability is used not only in the context of systems, but also in a more general sense.

Dignum [7] defines responsibility in the ART principles as a chain of responsibility that links the actions of the systems to all the decisions made by the stakeholders. We consider it to be the least accurately defined part of the ART principles, and thus have taken a more comprehensive approach to it in our research framework. According to the EAD, responsibility can be considered to be an attitude or a moral obligation [10].

Responsibility in the context of this study connects the designer to any stakeholders of the system. In order to be responsible, one must make weigh their own actions and to consciously evaluate their choices. A simplified way to approach responsibility is to ask "would I be fine with using my own system?".

To link this AI ethics discussion with SE practice, we have adopted the Commitment Net Model of Abrahamsson [1] to study AI Ethics in the context of Software Process Improvement (SPI). As we approach AI Ethics from the point of view of implementing it into practice in SE, we consider the utilization of extant theories in SPI useful for this purpose.

Developers' interests are driven by work-related concerns [1]. From the point of view of the developers, an important question to pose is: why would the developer act responsibly and take into account ethical issues? In order to understand commitment, we should first seek to understand the concerns of the developers which lead to actions, and together, form commitment. A task that may be perceived as time consuming, boring, or otherwise lacking in motivational elements, will still be executed because it plays a role in the developer's commitment behavior.

Commitment, accountability, responsibility and transparency can therefore be seen as a cycle with links (Figure 2). These links are explorative as little empirical data is currently available. We can hypothesize that by strengthening commitment towards the RESOLVEDD strategy, ethics will be implemented in the system through its use. Ethics, as defined by EAD, is made apparent through an increase in responsibility in design and the clarity of accountability, in order to help produce more transparency in AI development. Transparent project culture can likewise influence commitment, responsibility and accountability in design. In order to achieve this goal, the RESOLVEDD strategy should (1) support responsibility and responsible culture, (2) help developers to make more meaningful decisions in their own work, and (3) take into consideration ethical principles

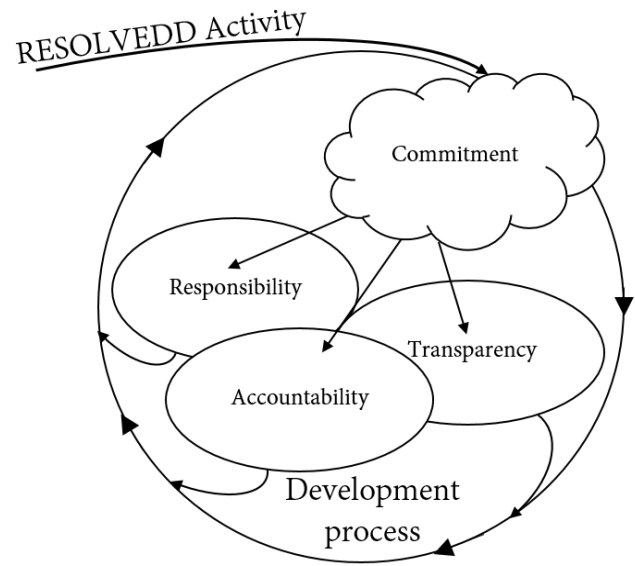

Figure 2 Research Framework for Ethically Aligned Design

such as accountability, privacy, autonomy, and fairness.

### 4. STUDY DESIGN

The RESOLVEDD strategy was empirically studied through a multiple case study. More specifically, we studied five student projects in which the RESOLVEDD strategy was utilized. Yin [27] explains that the use of multiple case study makes it possible to have multiple data sources with rich in-depth investigations that would not be possible with a survey. This approach also made it possible to analyze each case separately and to then validate the observations by cross-referencing.

The study was conducted in an Information Systems (IS) course at the University of Jyväskylä. Bachelor level students were introduced to the RESOLVEDD strategy as a part of system design and development methods. In the course, the students were given the task of developing a concept and prototype of a futuristic innovation that could be possible in the near future, but which was not considered currently plausible. The projects were carried out in five groups of 4-5 students. Choosing from a list, the students had to select one technology they would utilize as a part of their solution. For example, a team could choose to utilize Augmented Reality (AR) as a part of their solution in this fashion.

In the project, the use of the ethical method, RESOLVEDD, was given as one of the design requirements. The course spanned 10 weeks and consisted of eight weekly 5-6-hour workshop sessions and a project demonstration event held in the final week. During the workshop sessions, the students were introduced to the RESOLVEDD strategy in two lectures: 1) how to use the method, and 2) how to report their ethical considerations. The student were also given step-by-step instructions to the method and the project groups also had periodic RESOLVEDD strategy sessions with the teaching team where they had a chance to pose questions related to the method. At the end of the course, the teams presented their work in a project demonstration event. In the event they presented a demo of their solution and a poster where they had visualized the ethical issues, solutions to these issues and a justification to the actions taken in the design process.

Data for this study were collected by means of semi-structured interviews conducted after the course had concluded. The goal of the interviews was to (1) understand how the RESOLVEDD strategy had been used in practice in each project, and (2) how the ethical decision-making had been carried out in the projects, if at all. The interview questions were formulated based on the research framework. The semi-structured approach was applied to allow the respondents to elaborate on themes beyond the prepared questions. The interviews were conducted as group interviews with one project team at a time and recorded. The records were later transcribed, and the analysis was conducted using the transcripts.

Given the novelty of applying new ethical methods in AI ethics, and the current lack of existing literature related to our research questions, we adapted a qualitative approach, using open-ended interview questions. Moreover, we utilized a grounded theory inspired approach to analyzing them. We followed the recommendations of Heath and Cowley [15] in selecting a method that best suited our cognitive style and research environment. We utilized elements of the grounded theory approach proposed by Strauss and Corbin [24], aside from naming of the coding phases.

In practice, the transcripts were analyzed in the following manner. First, the transcripts were coded quote by quote and each quote was given a code describing its contents. Secondly, based on these codes, more abstract categories were introduced to group the individual quotes from each interview into general, re-occurring themes. Thirdly, this higher-level categorizing was validated by comparing the data from each interview. In this stage, we also sought to discover reoccurring themes across the five interview cases. From these reoccurring themes, core categories were formed and then compared to the research framework in order to determine how the principles of EAD were present in the projects (responsibility, meaningfulness, transparency and accountability), and what kind of commitment the developers exhibited towards implementing them. In discussing our findings, we present our key observations as Primary Empirical Conclusions (PECs).

## 5. FINDINGS

The findings from the analysis of the empirical data are reported here as topic-related Primary Empirical Conclusions (PEC). In total 5 PECs were formulated in the analysis. This section is structured into four sub-sections according to the research framework discussed in the preceding section. We illustrate some of our findings with relevant quotes from the respondents. However, our arguments are not solely based on the quotes but on our data in general.

### A. Commitment to Ethically Aligned Design

All five teams had rather critical sentiments towards dealing with ethical issues or using ethical tool as a part of their product design. Using an ethical tool was perceived as something completely novel to them, and they did not seemingly place value on considering the ethical aspects on their project. This was despite of the fact that the employed method is focused on helping its users detect ethical issues. When considering commitment to EAD, it is important to understand what the true concerns of the developers are. In this case, the teams were more concerned about the usefulness and viability of their product than its ethical aspects.

> "We don't want to do anything so absurd that it can't be actualized and that was probably our biggest motivator." -team 2

Aside from the usefulness and viability of their planned product, completing the projects on time and competing with the other teams were higher on teams' lists on concerns than ethics. The teams had difficulties seeing the ethical aspects as an activity that would help them to create better and more sustainable designs.

> "We spent time and effort on those tasks but it always felt very artificial because there was nothing to gain from it." -team 1

> "RESOLVEDD was a nice addition, but not absolutely necessary in this project. In another one it could be better." -team 4

The application of the RESOLVEDD strategy was part of the project requirements. Still, even after projects concluded, none of the teams thought that considering the ethical aspects of their product had been crucial to their success.

> "It [RESOLVEDD] was a burden for us. It was just there in the background, and we only remembered it was there when we had already designed something. We were not proactive with it." -team 2

Difficulties to develop concerns that would relate to ethics may also come from the nature of ethics itself. For the teams, ethics was something completely new. The educational system in IS studies directs the attention towards project requirements and other matters, and ethics are seldom discussed in relation to IS. Developing the ethical thinking of the students during the projects did not have the same kind of clear goals as the operational aspects of the project (e.g. were the requirements fulfilled). Similarly, some of the teams were frustrated that there were no "right" answers to the ethical issues that they faced:

> "At its best, an ethical tool would be tool that would inspire you to do good design. But RESOLVEDD didn't give us any answers to anything! If you put data into RESOLVEDD, you would not get anything out of it." -team 3

The teams also faced difficulties with RESOLVEDD. The teams were normatively committed to using RESOLVEDD to address the ethical issues faced in design. The normative commitment in this case was only externally enforced and thus not very strong.

> "Using RESOLVEDD felt forced since we didn't have that many ethical issues" - team 1

> "For us, the goal was not clear. We just needed to have some kind of product that supervisor would be ok with." - team 4

The teams did not consider RESOLVEDD helpful in reaching the project goals. Therefore, it was not considered useful by the teams. On the contrary, the teams considered it to be something that hindered their performance or drew their attention away from what they considered to be more important work. The teams did utilize it and reported their use of the tool, but only because it was required (= normative, external force). Notably, the teams remarked that the method needed to be adapted to better suit their context:

> "It [RESOLVEDD] felt like it didn't fit into our design process, so we had to adapt it, almost forcing it to work. So as an instrument it was not working." - team 3

> "For us it [RESOLVEDD] didn't work. We got much more out of having good conversations about ethical issues among the team. After those discussions, we just had to select some angle in order to force it into RESOLVEDD to get that requirement done." - team 2

The teams were, however, able to adapt successfully. They held group discussions where they discussed and addressed the ethical issues faced in their design processes. Thus, in practice, the teams used different methods to actually manage their ethical thinking. The RESOLVEDD strategy was then used to report their ethical thinking as a part of the course deliverables. None of the teams developed affective reasons to continue using the method after the projects concluded.

**PEC1:** While normative commitment to the use of Ethically Aligned Design brings immediate results, it will seize to exist when the external pressure is taken away. The RESOLVEDD strategy needs adaptation in application context. In practice, group discussions were seen effective in addressing the ethical issues.

### B. Transparency in design

Even though the teams were not affectively committed to using the ethical tool in their design process, they were required to follow the steps of the RESOLVEDD strategy and to produce documents that increased the transparency of the teams' decision-making processes. The teams adapted RESOLVEDD to fit their needs in order to carry out ethical analysis. The external pressure to use a specific method did not please the teams. Nonetheless, the necessitated use of the RESOLVEDD strategy method did increase transparency and ensured that the ethical discussions of the teams were documented for later use. The teams remained skeptical, however, whether their documentation would be beneficial.

> "Visualization of the RESOLVEDD-method seemed to be a waste of time and effort. Nobody would understand the drawn thing and all those lines in our picture." -team 3

The RESOLVEDD strategy primarily produced transparency in the design process itself rather than transparency in terms of the systems being designed. This may be in part due to the project setting where the focus was mostly on conceptualizing the product rather than the technical details. Furthermore, the developers were novices with little to no experience in AI development in practice. This may explain why the typical AI transparency issues, such as the black box thinking and understandability of the system actions, were omitted from the ethical considerations of the teams.

**PEC2:** When the RESOLVEDD strategy is followed step-by-step a paper trail is born where each decision made and the respective justification can be found. This produces transparency in the design process, but it does not promote transparency at the product layer.

### C. Accountability in design

The question of accountability divided the teams. It was not clear to the teams who could be held accountable for the design. Teams defended their position (not being accountable) by arguing that the systems are only concepts and prototypes. They outsourced the issue of accountability to the end user, or they were simply unable to explain how it would be managed from the legal or social viewpoints.

> "If this was a real life application, we would have had to think that if somebody steals the product and kills somebody with it, who would sue us? We didn't actively concern ourselves with studying any legal matters, we only considered those we realized by ourselves." -team 3

This all implies that the RESOLVEDD strategy did not support the idea of accountability or help the teams gather the needed knowledge for resolving the accountability issues.

**PEC3:** The RESOLVEDD strategy does not deliver accountability.

### D. Responsibility in design

Expecting the teams to engage in EAD and supporting their engagement in EAD by introducing an ethical tool made it possible to discuss ethical issues related to their current projects with the teams. However, our introduction to the RESOLVEDD strategy could have been better based on the data.

> "We did have a very independent and self-oriented group, but we knew that in the case of problems there would have been somebody there to help us. --- Then when RESOLVEDD came along, it was more like a nitpicking stuff. It wasn't very understandable." -team 4

In spite of the negative feelings expressed by the teams, reflecting on the ethical aspects became socially acceptable in the teams and in their development work. The developers shared their views on the responsibility issues among the team members in group discussions. These discussions activated reflections on the developers' own responsibility and raised the level of the developers' sense of responsibility.

> "We thought about the ownership issues and how it would be possible to misuse the [product]. Then we decided that it would be used as a vehicle and would be registered biometrically so no one else could use it." - team 3

> "We considered the loss of jobs and entire professions [resulting from AI]." - team 5

**PEC4**: Requiring Ethically Aligned Design activated reflections on the developers' own sense of responsibility

So far, we have established that the RESOLVEDD strategy promotes the use of EAD as described in PECs 2 and 4. However, we also found that the teams were not keen on using the method, nor were they satisfied with the results they obtained by doing so. External pressure for the use of the tool nonetheless created tangible results, promoted EAD, and even supported the developers' sense of responsibility. It remains an open question whether this is a merit to the RESOLVED strategy or whether this kind of improvement would have been achieved with any other ethical method as well.

**PEC5:** The mere presence of an ethical tool has an effect on ethical consideration creating more responsibility even when it the use of the method is not voluntary.

## 6. Discussion

On a general level, this study begins to bridge a gap discussed in existing literature. The IEEE guidelines for Ethically Aligned Design discuss a gap between research and practice in the area, underlining that work on the guidelines, as well as implementing AI ethics overall, has not carried over onto the field. In a similar vein, Morley et al. [21] note that the area is lacking in empirical studies actually testing the methods and tools that do exist. In this paper, we have begun to address these gaps by evaluating one ethical tool. Outside evaluating the specific tool, RESOLVEDD, our findings provide some insights into implementing AI ethics using any method or tool.

Indeed, PEC1 gives us some insights into commitment in the context of implementing ethics. By enforcing the use of an ethical tool top-down, it is possible to create normative commitment to implementing ethics (PEC4). This commitment, however, ceases to exist once the external pressure to utilize the tool ceases to exist. While this does support the implementation of ethics by making developers more responsible, if only while utilizing the tool, it does not result in any intrinsic motivation to implement ethics (PEC5).

This is interesting, however, as responsibility is typically considered to be intrinsically motivated and an attitude [10].

As for RESOLVEDD in particular [22], the tool supports one out of the two ethical principles that are currently considered to be the most important ones EAD: transparency and accountability. The use of RESOLVEDD produced transparency in the design process (PEC2). In utilizing it, the developers produced documentation on their decision-making, including reasoning behind their ethical choices as well as documenting alternate solution ideas that were ultimately discarded. Though transparency is considered required for accountability to be possible [10], RESOLVEDD did not produce accountability in the projects studied in this paper (PEC3). However, RESOLVEDD is not an ethical tool for AI ethics in particular, and thus does not account for the technical side of the system but only its overall design. It produces transparency of systems development (paper trail regarding decisions) but no transparency of data or algorithms.

Top-down adoption of ethical methods in organizations would seem to produce the wanted results, at least to some extent, and depending on the tool or method on question. Nonetheless, supporting the participatory adoption of such methods, as Morley et al. [21] suggest, would likely result in more ethical consideration from the developers. If they are intrinsically more motivated to implement ethics, they are arguably more likely to do so more meticulously.

On the other hand, adopting methods top-down is not a new proposition, especially in the context of SE. Many organizations made the move from waterfall to Agile development top-down after the management became convinced about the positive effects of Agile development, regardless of what the developers thought. While this induces change resistance, the developers will ultimately have to comply. Moreover, it can be difficult for developers to convince management, or even other developers, about the importance of ethics. Thus, while ethical methods should be designed with developers in mind, the point of entry into organizations for these methods may in fact be, e.g. the product manager.

Finally, the research framework formed in this study also has practical implications by making the level of Ethically Aligned Design evaluable. We have shown, initially, that while it is possible to introduce EAD by force, results will not sustain over time. The RESOLVEDD strategy needs to be adjusted in practice. One important adjustment done by our case teams was the introduction of group discussions as the primary means to do EAD in practice. Thus, a possible avenue for tailoring is to identify what are the practices that actually lead to favorable outcomes increasing transparency, responsibility and accountability.

### A. Limitations of the Study

The primary potential limitation of this study are its sample size and the use of student projects. However, in relation to using students as subjects for data collection, Höst et al. [16] argued that the differences between students and professionals in SE is minor and not statistically significant. In fact, they recommend the use of students in SE studies. Runeson [23] found similar improvement trends between undergraduate, graduate and professional study groups. For a novel topic in the field (such as EAD here), the students provide an excellent platform for an empirical evaluation, method development and experimentation.

Additionally, in relation to our sample size, we acknowledge that five projects is not a large sample. Nonetheless, Eisenhardt [9] note that 4 to 10 cases typically work well in case study research, outside particularly in-depth case studies, which may utilize fewer cases. They also highlight the suitability of case studies for novel research areas [9]. While AI ethics is not a novel area as such, empirical studies in the area are lacking, especially in relation to methods.

### 7. CONCLUSIONS & FUTURE WORK

In this study, we have evaluated the RESOLVEDD strategy for ethical decision-making through an exploratory, multiple case study of five student projects. The main results of this study are as follows: (1) While normative commitment to the use of Ethically Aligned Design brings immediate results, it will cease to exist when the external pressure is taken away. (2) An ethical method (RESOLVEDD) that necessitated tracking the decisions that were made produced transparency in the design process. (3) The RESOLVEDD strategy does not deliver accountability. (4) Requiring Ethically Aligned Design from the developers also resulted in responsibility in the developers. (5) The mere presence of an ethical tool has an effect on the ethical consideration exerted by developers, creating more responsibility even when the use of the method is not voluntary.

Thus, forcefully implementing an ethical tool or method can further the implementation of ethics. A top-down approach to introducing a tool or method for implementing ethics can serve as a starting point for ethical development in an organization. However, normative commitment does not seem to result in any intrinsic motivation to implement ethics among developers. I.e. this does not motivate the developers to implement ethics out of their own volition.

Based on these results, the following theoretical implications can be made. The formed research framework where ethical principles are combined with concept of commitment is a functional approach for evaluating the inclusion of ethics in design. Understanding the mechanics related to the developers' commitment(s) has a crucial role in furthering the inclusion of ethics in design.